# Antiferromagnetic properties of a water-vapor-inserted YBa$_2$Cu$_3$O$_{6.5}$ compound studied by NMR, NQR, and $\mu$SR

A. V. Dooglav, A. V. Egorov, I. R. Mukhamedshin, and A. V. Savinkov
*Magnetic Resonance Laboratory, Kazan State University, 420008 Kazan, Russia*

H. Alloul, J. Bobroff, W. A. MacFarlane, and P. Mendels
*Laboratoire de Physique des Solides, Universite de Paris-Sud, 91405 Orsay, France*

G. Collin, N. Blanchard, and P. G. Picard
*Laboratoire de Leon Brillouin, CE Saclay, CEA-CNRS, 91191 Gif-sur-Yvette, France*

P. J. C. King and J. Lord
*ISIS, Rutherford Appleton Laboratories, Chilton Didcot, OX11 OQX, England*

We present a detailed NQR, nuclear magnetic resonance (NMR), and $\mu$SR study of the magnetic phase obtained during a topotactic chemical reaction of YBa$_2$Cu$_3$O$_{6.5}$ high-temperature superconductor with low-pressure water vapor. $^{65}$Cu-enriched samples have been used for NQR/NMR studies which allows to get a good resolution in the Cu(1) NQR and Cu(2) zero field NMR (ZFNMR) spectra. It is shown that the NQR spectrum of the starting material transforms progressively under insertion of water, and almost completely disappears when about one H$_2$O molecule is inserted per unit cell. Similarly, a $^{65}$Cu ZFNMR signal characteristic of this water inserted material appears and grows with increasing water content, which indicates that the products of the reaction are nonsuperconducting antiferromagnetic phases in which the copper electronic magnetic moments in the CuO$_2$ bilayers are ordered. The use of $^{65}$Cu-enriched samples allowed us to reliably resolve three different copper resonances which correspond to different internal magnetic fields. The antiferromagnetic phases are also felt by proton NMR which reveals two sites with static internal fields of 150 and about 15 Gauss, respectively. $\mu$SR studies performed on a series of samples prepared in the same way as the $^{65}$Cu-enriched ones reveal two muon sites with the same local fields as the proton sites, which vanish at $T \approx 400$ K. This indicates that muons preferentially occupy proton vacancy sites, and that the magnetic phases have similar Néel temperatures as the other bilayer undoped cuprate compounds. An analysis of the internal fields on the different spin probes suggests that they can be all assigned to a single magnetic phase at large water content in which the Cu(1) electron spins order with those of the Cu(2). The detailed evolution of the spectra with the progressive increase of water content is shown to be compatible with a coexistence of phases during the early stages ot the reaction. It appears that even samples packed in paraffin underwent a transformation of a substantial part of the sample after 6 years storage in atmosphere. Samples packed in Stycast epoxy resin heated moderately to a temperature (200°C) undergo a reaction with a coexistence of epoxy decomposition products which yield the formation of the same final compound. It is clear that such effects should be considered quite seriously and avoided in experiments attempting to resolve tiny effects in these materials, such as those performed in some recent neutron scattering experiments.

## I. INTRODUCTION

The underdoped regime of high T$_c$ cuprates is still a matter of intense investigation. Many questions about the origin of the pseudogap remain. It is often proposed that the broken symmetry in the superconducting state is complex and might combine a superconducting order parameter and an orbital or magnetic order parameter. Experimental investigation on such kinds of order typically have to deal with weak signatures which, for example, require high sensitivity neutron scattering experiments.[1,2] Such studies have been undertaken on the so called ortho II phase of YBa$_2$Cu$_3$O$_{6+x}$ which is stabilized for $x=0.5$. We had undertaken systematic Cu NMR experiments in zero field (ZFNMR) on this compound in which we had found sizable NMR signals in an internal field, indicating the existence of an antiferromagnetic (AF) order.[3] The spectroscopic features of these signals were totally different from those observed in the AF parent compound YBa$_2$Cu$_3$O$_6$ and had to be associated with a different magnetic phase. However, these signals were found to occur mostly in samples aged for a long time in air and appeared to result from a reaction with air moisture.[4] These phases are, therefore, serious potential contaminants of any experiment aiming at studying magnetic responses of these materials. Our preliminary results certainly have led some groups dealing with neutron scattering to be concerned with this problem.[5] However, we did not up to now characterize in detail the magnetic properties of these AF phases.

A. V. DOOGLAV *et al.*

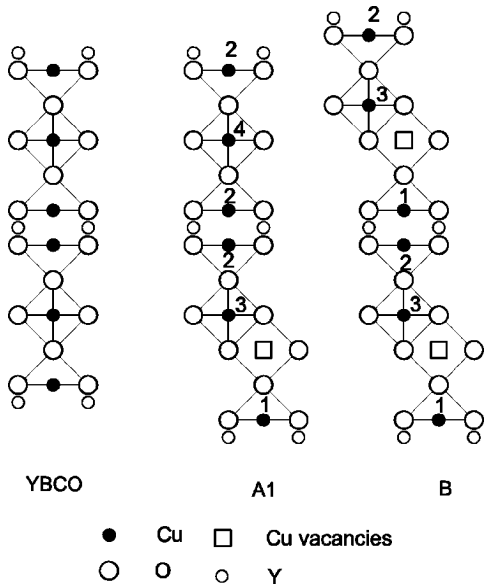

FIG. 1. Structures of the initial phase, intermediate phase $A1$ and the final phase $B$ (from Ref. 8). The structurally equivalent copper sites are labeled with the same numbers.

We found out that intensive investigations on the degradation of high-temperature superconductors by water vapor have been performed since the very beginning of the HTSC story. Attention was mainly focused on the loss of superconductivity due to water. Some papers reported that even packing the HTSC powder in epoxy[6] or storing it in a desiccator[7] did not allow one to completely avoid the reaction with water, so that a small amount of the degraded phases is likely to be present in practically all samples studied.

Most of the early papers on the reaction of 123 with water reported the appearance of compounds such as $Y_2BaCuO_5$ ("green phase"), $CuO$, $BaCO_3$, etc., which are the final decomposition products, and display well known properties. The more recent studies (see Ref. 8 and references therein) done by high-resolution electron microscopy, thermogravimetry and x-ray investigations have shown that the 123 superconductors with $x < 0.9$ react with water vapor in a topotactic mechanism. The reaction occurs in two stages in the bulk of the material in the rather narrow temperature range of $75-250\,°C$ (for $YBa_2Cu_3O_{6.5}$). At small water uptake (up to 0.5 water molecule per $YBa_2Cu_3O_{6.5}$ unit cell), performed at temperatures $T < 180\,°C$, an intermediate phase $A1$ is formed (Fig. 1) which has a unit cell parameter $c = 5.029$ nm. The authors of Ref. 8 suggest that this phase has a $T_c$ slightly higher than that of the starting material. Upon heating, this phase is transformed in a highly disordered $A$ phase (not described well enough in Ref. 8 to be shown in Fig. 1) with $c = 2.64$ nm which quickly transforms at room temperature in the presence of water vapor into a nonsuperconducting $B$ phase with $c = 2.72$ nm. This final product of the reaction is the so-called pseudo-1248 phase $H_{2Z}YBa_2Cu_3O_{6+x+z}$, isostructural with the familiar 1248 superconducting compound but with 50% of $Cu(1)$ sites vacant. The fact that it can be considered as a "parent" nonsuperconducting compound of the 1248 family (which normally does not exist) has motivated us to perform a systematic study of the changes of electronic properties occurring in the process of conversion of the 123 superconductor into the pseudo-1248 phase upon the insertion of water.

Although this paper addresses a problem already considered in our preliminary publications,[4,9] here, we study carefully the evolution of various spectral information as a function of water content in order to attempt to understand the electronic properties and magnetic structure of the phases observed upon hydration of $YBa_2Cu_3O_{6+x}$ compounds. Measurements performed on a $^{65}Cu$-enriched sample allowed us to follow the disappearance of the NQR spectrum with water content $Z$, to separate clearly three different components in the $^{65}Cu$ ZFNMR spectra, and to follow their evolution with the water content $Z$.

The paper is organized as follows. In Sec. II we give details of the preparation of the samples and of water insertion in our $YBa_2Cu_3O_{6+x}$ starting materials. Elementary sample characterization was performed and allowed us to demonstrate that the superconducting fraction is progressively suppressed with water insertion, and that superconductivity disappears completely when one molecule of $H_2O$ is inserted per unit cell. In Sec. III we recall the characteristics of the NQR of the starting $YBa_2Cu_3O_{6.5}$ material and of the ZFNMR of the magnetic phase of the parent compound $YBa_2Cu_3O_6$, which are essential for the understanding of the water insertion process. In Sec. IV the NQR of the water-$YBa_2Cu_3O_{6.5}$ samples allows one to follow the progressive disappearance of the metallic phases. Similarly the Cu ZFNMR spectra taken on the $^{65}Cu$ enriched sample allow us to detect three Cu signals with different internal fields and quadrupole frequencies. The modification of the spectra with increasing water content are studied in great detail. Finally, $\mu$SR spectra reveal the existence of two muon sites which sense two distinct internal fields occurring in the magnetic phases and which disappear at a similar Néel temperature. Two internal fields are detected as well on the proton sites, and have the same values as those found on the muon sites. In Sec. V all these results are analyzed together and allow an assignment of the Cu ZFNMR to the actual sites of the final product determined by x-ray scattering. The combination of all the information from the different techniques seems to converge towards a solution for which both $Cu(2)$ and $Cu(1)$ moments order magnetically in the water saturated phase. Intermediate products of the reaction appear AF as well, with minor changes in the magnetic spectral features. Since epoxy resin is thought to protect the sample from water, a few samples packed in epoxy were also studied. A brief description and analysis of the data on such samples are given at the end of Sec. V.

## II. SAMPLE CHARACTERIZATION

### A. Sample preparation

We have synthesized a series of $YBa_2Cu_3O_{6+x}$ samples with $x$ ranging from 0.4 to 1 using conventional solid state reaction of powdered $Y_2O_3$, BaO and CuO at about $940\,°C$, interrupted periodically for grinding. These samples with a particle size of the order of 30 $\mu$m were used in experiments both as free powders and packed in Stycast 1266A epoxy resin. To get better resolution in copper NQR and ZFNMR spectra a sample of $YBa_2Cu_3O_{6.5}$ enriched with the $^{65}Cu$ isotope (99.6%) was also synthesized. Although the results for all samples were qualitatively similar, systematic measurements were done on free powders of the $^{65}Cu$-enriched sample $YBa_2Cu_3O_{6.5}$, and on two samples with natural copper content, $YBa_2Cu_3O_{6.5}$($T_c$=56 K) and $YBa_2Cu_3O_{6.42}$($T_c$=20 K).

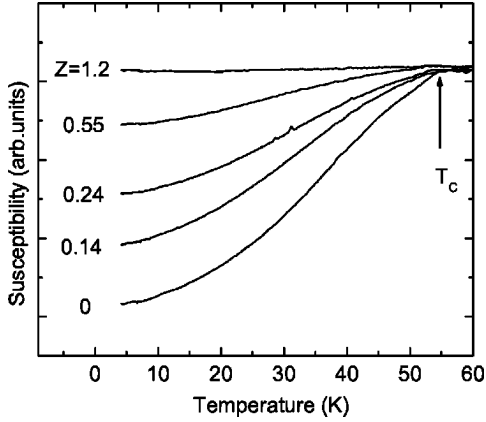

FIG. 2. Temperature dependence of the susceptibility for $YBa_2Cu_3O_{6.5}$ with different water uptake $Z$, showing that $T_c$ is not modified and that the superconducting fraction diminishes.

The free powders were annealed at 110°C in a closed furnace in air with a vapor pressure (70–80 mbar) created by evaporating an adequate amount of water. The water uptake was determined by the weight increase of the samples. We have found that the uptake rate is strongly modified if water molecules have been absorbed on the powder surface in air at room temperature. In the conditions mentioned above and when the samples have been preliminarily dried in vacuum, the water uptake $Z$ increases almost linearly in time with a rate $\Delta Z/\Delta t \approx 0.05$ per day. A series of $YBa_2Cu_3O_{6.5}(H_2O)_Z$ and $YBa_2Cu_3O_{6.42}(H_2O)_Z$ were produced with nominal $Z$ varying from 0.03 to 1.2.

### B. X rays

The x-ray diffraction spectrum taken for the sample with a maximum water uptake $Z$=1.2 confirm that the main final product of the reaction is as for Ref. 8 the tetragonal phase with $c$=2.72 nm, i.e., the pseudo-248 structure with 50% defects on chain copper positions (Fig. 1, phase $B$). Definite signs of small residual contamination by the $A1$($c$ =5.03 nm), the $A$($c$=2.64 nm) and the starting 123 material ($c$=1.17 nm) phases were present in the x-rays data taken on the $Z$=1.2 sample.

### C. Susceptibility data

The temperature dependence of the susceptibility for the samples with different water uptake are shown in Fig. 2. It is clearly seen that the superconducting volume fraction decreases with water uptake while $T_c$ remains practically unchanged. The sample with a maximum water uptake is com-pletely nonsuperconducting. This indicates that the reaction with water vapor creates new non superconducting phases, which progressively replace the starting phase until no superconducting volume remains. We could not sense in these susceptibility data any significant increase of the $T_c$ onset, which would be attributed to the $A1$ phase. So, if this phase is present in our samples during the intermediate stages, it does not appear to us to correspond to a higher $T_c$. As we shall see later it might be magnetic as well.

## III. COPPER NQR AND ZFNMR SPECTRA IN $YBA_2CU_3O_6$ AND $YBA_2CU_3O_{6.5}$

Before describing the actual data in the present $YBa_2Cu_3O_{6.5}$- water samples, we recall briefly as a reference the well known results in the parent $YBa_2Cu_3O_6$ AF compounds from ZFNMR of the planar Cu(2) site and by NQR of the intercalated Cu(1)3d$^{10}$ site. Also we recall the NQR signals of both sites obtained in the $YBa_2Cu_3O_{6.5}$ metallic phase, free of any water contamination. This then allows to follow the evolution of the spectra detected with increasing water content.

### A. Copper ZFNMR and NQR in antiferromagnetic $YBa_2Cu_3O_6$

At low oxygen content ($x$=0–0.2) the $YBa_2Cu_3O_{6+x}$ compound is an insulating antiferromagnet with a Néel temperature of about 410 K for $0 \leqslant x \leqslant 0.1$. The electronic magnetic moments of $Cu^{2+}$(2) ions are antiferromagnetically ordered both within the bilayers and between the bilayers (AFI phase) and form a quasi-2D antiferromagnetic network. Cu(2) nuclei experience a large hyperfine magnetic field produced by the on-site and four neighboring $Cu^{2+}$ ordered electronic magnetic moments $\mu_{eff} \approx 0.65 \mu_B$. The nuclear magnetic resonance (NMR) of Cu(2) nuclei observed in this internal magnetic field $\mathbf{H}_{int}$ is usually referred to as Cu(2) zero field NMR (ZFNMR), since one does not need to apply the external magnetic field to observe the Cu(2) NMR.

The Cu(2) ZFNMR spectrum is described by the nuclear spin Hamiltonian[10]

$$\mathcal{H} = -\gamma_n \hbar \mathbf{H}_{int} \cdot \mathbf{I} + \frac{eQV_{zz}}{4I(2I-1)} \left\{ 3I_z^2 - I(I+1) + \frac{1}{2}\eta(I_+^2 + I_-^2) \right\},$$ (1)

where $\gamma_n$ is the nuclear gyromagnetic ratio ($\gamma_n/2\pi$=11.28 and 12.09 MHz/T for $^{63}Cu$ and $^{65}Cu$ nuclei, respectively), the $x,y,z$ axes are the principal axes of the crystal electric field gradient (EFG) tensor chosen so that $|V_{zz}|>|V_{xx}|>|V_{yy}|$, $\eta=|V_{xx}-V_{yy}|/V_{zz}$ the asymmetry parameter showing the deviation of the EFG symmetry from axial. Both naturally abundant copper isotopes, $^{63}Cu$ (69.1%) and $^{65}Cu$ (30.9%), have a nuclear spin $I$=3/2, with different quadrupolar moments, $^{63}Q$ and $^{65}Q$. The on-site $A$ and near neighbor $B$ hyperfine couplings produce a total magnetic field $H_{int}=|A-4B|\cdot\mu_{eff}\approx80$ kG at the Cu(2) nucleus. Thus the ZFNMR spectra of $^{63}Cu$(2) and $^{65}Cu$(2), observed in the frequency range 70–110 MHz, consist of three lines for each copper isotope. They are characterized by $\eta \approx 0$, with the principal axis of EFG parallel to the crystal *c*-axis, and $\mathbf{H}_{int} \perp c$.[11] Quite generally the appearance of a copper ZFNMR in the frequency range 70–180 MHz in a compound signals the occurrence of long-range antiferromagnetic ordering of the $Cu^{2+}$ moments.

In the AF $YBa_2Cu_3O_{6+x}$ compounds the nuclei of diamagnetic $Cu^+(1)$ ions only sense small (about 1 kG) hyperfine magnetic fields $H_{int}$ transferred from the two adjacent $Cu^{2+}(2)$ ions. In the AFI phase these Cu(2) electronic magnetic moments are antiparallel, and their contributions to $H_{int}$ cancel. So, for Cu(1) nuclei one observes a pure NQR spectrum described by the second term of Hamiltonian (1). This observation is in fact also a direct evidence that the spins on the $Cu^{2+}$ are indeed AFM ordered between bilayers. For each Cu isotope, the NQR spectrum consists of a single line at a frequency[10]

$$\nu_Q = \frac{eQV_{zz}}{2h}\sqrt{1 + \frac{1}{3}\eta^2}, \qquad (2)$$

which is equal to 30.2 MHz for $^{63}Cu(1)$ in $YBa_2Cu_3O_6$. The NQR line for $^{65}Cu$ is observed at a frequency $^{63}Q/^{65}Q$ = 1.081 times smaller than for $^{63}Cu$.

The AF order can be modified if some trivalent or magnetic ions are inserted in the chains of $YBa_2Cu_3O_6$, for example, $Al^{3+}$ when $Al_2O_3$ crucibles are used for the synthesis, or upon intentional substitution of Ni on the chain sites.[12] Above a critical impurity concentration a ferromagnetic ordering of the Cu(2) magnetic moments neighboring the Cu(1) occurs at low *T*. The hyperfine magnetic fields from the two adjacent $Cu^{2+}(2)$ ions then add at the Cu(1) nuclei, which lifts the degeneracy of the Cu(1) pure NQR levels. In this so-called AFII phase the corresponding splitting of the NQR line allows to determine $H_{int} = 2$ kG.[12]

## B. Copper NQR in superconducting $YBa_2Cu_3O_{6.5}$

Doping the chains with oxygen increases the concentration of mobile holes in the $CuO_2$ planes, so that for *x* $\geq 0.35$ the long-range antiferromagnetic order is lost, the internal magnetic field at Cu(2) sites averages to zero, and one can then observe the pure NQR spectra of Cu(2). This oxygen insertion in the chains also converts neighboring $Cu^+$ into $Cu^{2+}$. The disruption of the $Cu^{2+}$ chains by $Cu^+$, and/or mobile holes left in the chains, prevent any static ordering of the $Cu^{2+}(1)$ electronic magnetic moments. Hence, the absence of static internal field at the Cu(1) nuclei again allows the observation of pure NQR spectra of Cu(1).

The NQR frequencies of both Cu(2) and Cu(1) in $YBa_2Cu_3O_{6+x}$ are well known for all oxygen contents.[13] For *x* $\approx 0.5$, the case of interest here, the observed copper NQR spectra are quite intricate as the oxygen order is not perfect in the chains in real samples, so that diverse copper sites occur with different oxygen coordination. For simplicity we shall discuss here the $^{65}Cu$ copper NQR spectrum of the $^{65}Cu$-enriched sample (Fig. 3). The corresponding lines for $^{63}Cu$ in natural copper samples are located at frequencies 1.081 times higher than those for $^{65}Cu$. Since the nuclear spin-lattice relaxation rate of the "plane" copper is in general significantly faster than that of "chain" copper,[3] it is possible to separate their spectra [Figs. 3(b) and 3(c)]. The large difference of their transverse relaxation rates can be used as well to distinguish the two components. The fast relaxing part of the $^{65}Cu$ NQR spectrum beyond 24 MHz [Fig. 3(c)] is assigned to the plane copper nuclei, and consists of two lines at 28.3 and 25.7 MHz. The corresponding sites have different oxygen coordination in the neighboring chain.

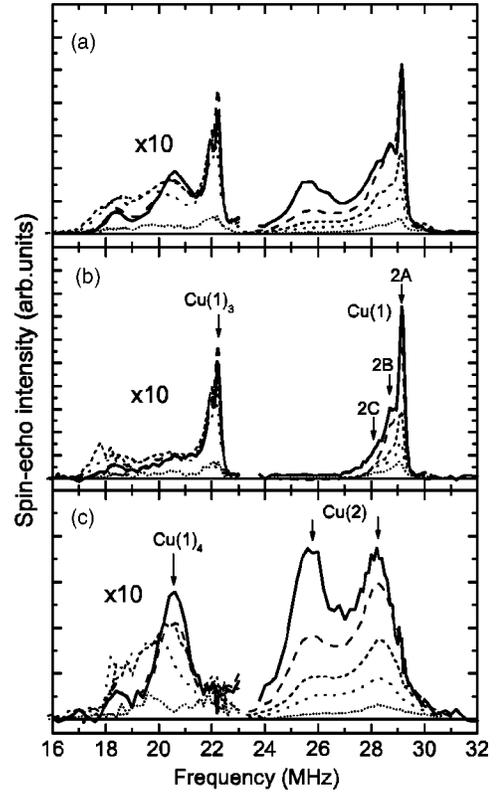

FIG. 3. Copper NQR spectra for the $^{65}Cu$-enriched $YBa_2Cu_3O_{6.5}$ sample with different water uptake Z: (a) Nonsaturated spectrum taken with a 1s repetition time. (b) was obtained as the difference of spectrum (a) and of a scaled spectrum with short repetition time (20 ms), which emphasizes then the slow-relaxing part of the spectrum [dominated by the Cu(1)]. (c) was obtained as the difference of the spectrum with short repetition time (20 ms) and a scaled spectrum (a), which emphasizes then the fast-relaxing part of the spectrum (dominated by the Cu(2)). Solid lines—as-prepared sample; dashes, short dashes, dots, short dots—$Z$=0.03, 0.38, 0.59, and 0.93, respectively. The spectra are 10 times enhanced in the frequency range below 23 MHz. See the text for details on the assignments of the various signals.

The assignment of the Cu(1) NQR lines is done in Ref. 14. The main ordered phase for *x* $\approx 0.5$ only contains two Cu(1) sites located either in "full" ...-Cu-O-Cu-... chains between "empty" ...-Cu-Cu-... chains, or conversely. The former is a fourfold coordinated copper which corresponds to the fast relaxing line at 20.8 MHz, typical of the Cu(1) site in the $YBa_2Cu_3O_7$ Ortho I compound. The second twofold coordinated copper labeled as $Cu(1)_{2A}$ in Ref. 14 corresponds to the sharp slow relaxing $^{65}Cu$ NQR line at 29.15 MHz. Other twofold coordinated copper give resolved lines depending on the occupancy of the neighboring chains. The line at 28.7 MHz is from copper Cu(1)$_{2B}$ of empty chains of small islands composed of one full and two empty chains. The line at 27.8 MHz corresponds to Cu(1)$_{2C}$ with no oxygen in the chains, typical of YBa$_2$Cu$_3$O$_6$. Finally, the slow relaxing sharp $^{65}$Cu line at 22.2 MHz is ascribed to the threefold coordinated Cu(1)$_3$ at the ends of disrupted Cu-O chains.

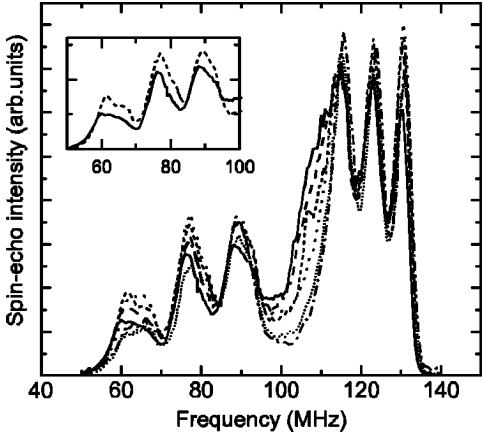

FIG. 4. Copper ZFNMR spectra for YBa$_2$Cu$_3$O$_{6.5}$ synthesized with enriched $^{65}$Cu with different water uptake $Z$. Solid, dashed, short dashed, dotted, short dotted, and dash-dotted lines correspond to $Z$=0.93, 0.77, 0.59, 0.38, 0.185, and 0.085, respectively. The amplitudes of the spectra are normalized by the water uptake $Z$. The fact that the low-frequency spectrum slightly shifts to lower frequency with water uptake is emphasized in the inset.

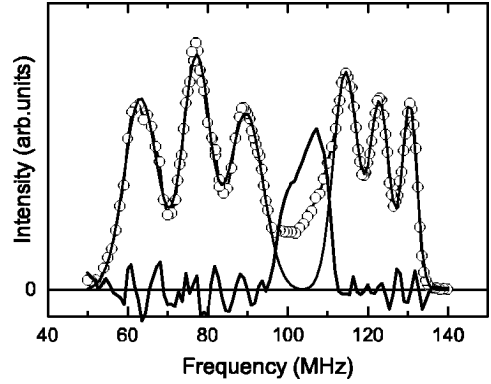

FIG. 5. Frequency corrected copper ZFNMR spectrum for $^{65}$Cu enriched YBa$_2$Cu$_3$O$_{6.5}$(H$_2$O)$_{0.38}$. The sharp lines of the lower frequency and higher frequency parts of the spectrum are fitted with six Gaussians. The central part of the spectrum ( magnified by a factor 2 in the figure) is obtained after subtraction of the fitting curve from the experimental spectrum.

## IV. WATERED SAMPLES

### A. Copper NQR

As was already found earlier for samples with natural copper isotopes,[4] the copper NQR spectrum intensity decreases rather inhomogeneously when water uptake increases. The overlap between the $^{65}$Cu and $^{63}$Cu spectra does not take place anymore in the $^{65}$Cu-enriched sample so that we could study in some detail the evolution of the copper NQR spectrum (Fig. 3) with water uptake. The spectrum for the as-prepared powder sample is shown by a thick line in Fig. 3. After keeping the sample for two weeks in ambient atmosphere at room $T$ we found that the intensity of the 25.7 MHz Cu(2) line was slightly reduced. We attribute it to a very small water uptake from ambient atmosphere.

In general, the results obtained for the $^{65}$Cu enriched samples confirm our previous findings, but a more homogeneous disappearance of the copper NQR spectrum with water uptake was observed. This is likely due to the higher water vapor pressure used for loading the $^{65}$Cu-enriched sample with water as compared to our previous series of samples.[4] Anyhow, the narrow NQR line of the Cu(1)$_{2A}$ nuclei belonging to the empty …-Cu-Cu-… chains and the Cu(2) line at 25.7 MHz disappear faster with water uptake than the other parts of the NQR spectrum. This confirms the conclusion that the insertion of water molecules proceeds through the empty

Cu(1) chains. The broad fast relaxing Cu(2) NQR line at 25.7 MHz might then be assigned to Cu(2) nuclei belonging to the well-developed OrthoII phase, most likely to those located just above or below the empty Cu(1) chains. The intensity of the Cu(1)$_3$ NQR line does not change significantly at the beginning of the reaction and disappears only at rather high water uptake. A similar behavior can be noticed also for the Cu(1)$_4$ NQR line, which confirms that water diffusion is hampered in areas of oxygen-filled chains within the crystallites.

### B. Copper ZFNMR in the ordered magnetic phases

Here, we study the ZFNMR of Cu which gives information on the appearance of ordered magnetic phases. To get a better sensitivity, all measurements were done at $T$=4.2 K.

The copper ZFNMR spectra of free-powder annealed samples are shown in Figs. 4 and 5. As-prepared samples did not exhibit any traces of copper ZFNMR spectra, but after two weeks in ambient atmosphere, distinct spin echo signals were detected at copper ZFNMR frequencies. The signal intensities were, however, too small to allow us to record a full spectrum, but this confirms that water contamination starts readily at room $T$. Even minute amount of water reacting with the 123 compound induces the appearance of magnetic phases altogether with the changes in the copper NQR spectrum mentioned in the previous Section.

The spectrum consists of three groups of lines which, contrary to the data published earlier,[4] are well resolved in the $^{65}$Cu-enriched sample (Figs. 4 and 5). The low-frequency (58–95 MHz) and high-frequency (110–132 MHz) groups both consist of three lines and resemble the Cu(2) ZFNMR spectrum in YBa$_2$Cu$_3$O$_6$. From the frequencies of the three $^{65}$Cu ZFNMR lines, it is impossible to extract the four parameters of Hamiltonian (1), that is $H_{int}$, $V_{zz}$, $\eta$, and the angle $\Theta$ between the internal magnetic field and the principal axis of the EFG. The only parameter which could be reliably estimated is $H_{int}$, as the frequency of the central $m$= $-1/2 \leftrightarrow +1/2$ transition is determined mainly by $H_{int}$ and depends on the quadrupolar effects only through second order perturbation.[10] If we assume further that $\eta=0$ and $\Theta=90°$, as for Cu(2) in $YBa_2Cu_3O_6$, one can fit both spectra fairly well with $H_{int}=62$ kG, $\nu_Q=26$ MHz for the low-frequency and $H_{int}=101$ kG, $\nu_Q=16$ MHz for the high-frequency one.

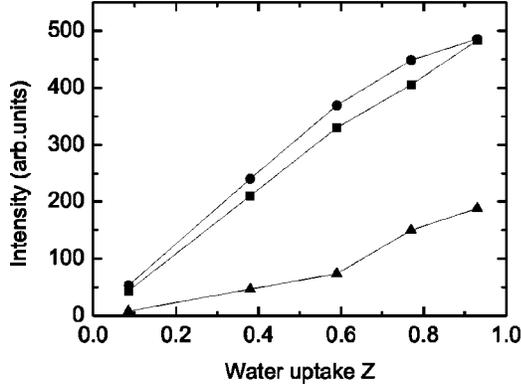

FIG. 6. Dependence on water uptake $Z$ of the intensities of different parts of the copper ZFNMR spectra (after frequency and $T_2$ correction). Circles, squares and triangles correspond to the intensities of the low-frequency, high-frequency and central part of the spectra, respectively.

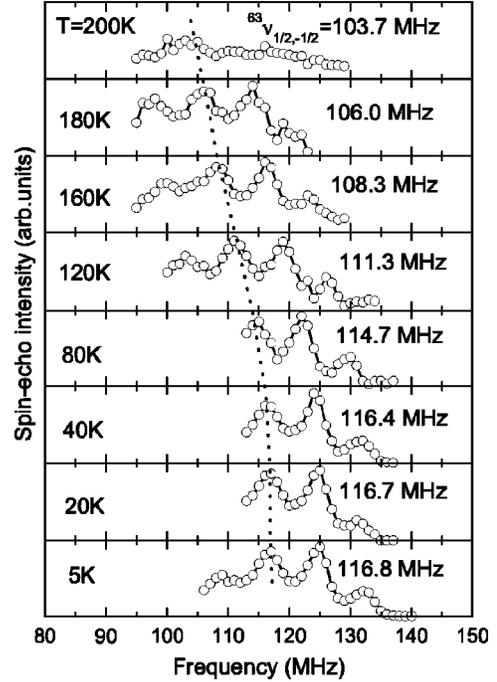

FIG. 7. Temperature dependence of the high-frequency part of copper ZFNMR spectrum in a sample with natural copper. The dashed line follows the temperature variation of the high-frequency central transition for [63]Cu. Its corresponding frequencies are given on the right hand side of the spectra.

The above mentioned assumptions $\eta=0$ and $\Theta=90°$ do not hold for the third, central part of the spectrum which does not exhibit three, but only two poorly resolved lines. If we assume $\eta=0$, the spectrum might be fitted with $H_{int}=85$ kG, $\Theta=55°$. If, on the contrary, we assume $\Theta=90°$, one gets $\eta\approx0.5$ and $H_{int}=85$ kG. In any case, the central spectrum is very different from the two others, and corresponds to a value of $H_{int}$ which is rather insensitive to the choice of $\Theta$ and $\eta$.

The intensities of the three spectral groups scale linearly with water uptake for small $Z$ (Fig. 4), while for $Z\geqslant0.5$ the component at 96–112 MHz begins to grow faster than others and reaches its maximal intensity in the sample with maximum water uptake. To perform a more quantitative analysis, we have corrected the spectra of Fig. 4 by extrapolating the spin echo intensity to zero time delay and determined the signal intensity as $\int(f(\omega)/\omega^2)d\omega$ for the three components of the spectra. The variation with water uptake of their intensities is reported in Fig. 6. It can be seen that the evolution with water uptake for the two outer spectra is identical which suggests that they pertain to the same phase, which grows regularly with water content. They apparently correspond as well to a similar number of Cu nuclei. The central part of the spectra corresponds to less nuclei than the two others and increases faster at a large water content.

At high temperature, the spectra can no longer be taken with good accuracy, but we could however follow the high-frequency part up to 200 K. It can be seen in Fig. 7 that the frequency of the central transition of the spectra, which is determined mainly by the internal field, decreases with temperature, as expected for an AF phase.

### C. Zero field $\mu$SR

In order to get a better insight into the nature and the $T$-evolution of the AF phases detected by NMR, we per-formed $\mu$SR studies on a series of compositions in $YBa_2Cu_3O_{6.4}(H_2O)_Z$, with $Z=0.62$, 0.9, and maximum 1.2 water content up to high temperatures where the ZFNMR is not observable. In contrast to a previous report[9] *all* our samples here were treated by the same process, as described in the sample preparation section. A *unique* behavior, as described below, is observed in all our samples at variance with our previous preliminary publication.[9] We ascribe this to the way our $Z=0.64$ and 0.2 samples had been produced, i.e., by annealing the $Z=0.9$ batch in vacuum. For those samples, we suspect some sample decomposition occurred while annealing, which leads to a signal similar to that in $YBa_2Cu_3O_6$ with a typical frequency of 4 MHz.[15] The data presented here were taken at the Paul Scherrer Institut (Switzerland) and at ISIS (England) facilities using conventional zero field $\mu$SR and weak transverse field techniques as described below.

Aside from a weak high frequency signal ($\sim$18 MHz at $T\rightarrow0$) analogous to that attributed to a small fraction of muons sitting near the AF ordered $CuO_2$ planes in $YBa_2Cu_3O_{6+x}$,[15] ZF $\mu$SR data taken in the range 15–420 K reveal a single damped oscillating signal with a characteristic frequency around 2 MHz at $T\rightarrow0$, as can be seen from the representative asymmetry spectra for $Z=1.2$ and $Z=0.62$ displayed in Fig. 8. This, hereafter named "2 MHz" signal, is the major $\mu$SR signature of the existence of an internal field of the order of 150 G, hence of the magnetic ordering in our samples. Moreover, for $Z=1.2$, we could clearly observe an additional large Kubo-Toyabe (KT) component displaying a characteristic dip, corresponding to muons feeling a small and distributed static field ~10 G. We will show hereafter that these two signals can be associated with a single AF phase and two muon sites, likely located near O(4) and O(1) for the 2 MHz and KT signals, respectively. Analogous data can be obtained using proton NMR up to 420 K and the locations of the muon and the proton will be discussed further on, in the analysis section of the paper.

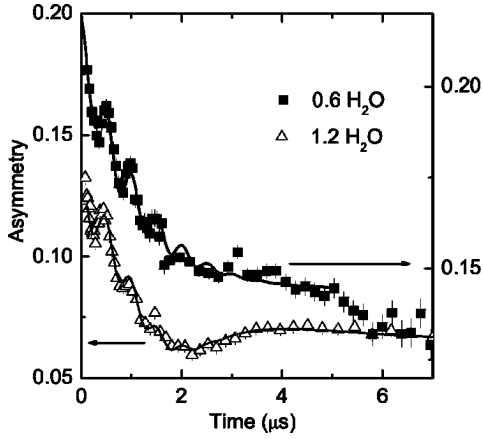

FIG. 8. Plot of the muon asymmetry versus time for the $Z$ =0.6 and 1.2 samples. A 2 MHz component is evident for both samples. Notice the "dip" in the asymmetry of the $Z$=1.2 sample which is characteristic of a Kubo-Toyabe additional component.

The data for $Z$=1.2 was extensively fitted, using the sum of a Gaussian damped oscillating term (2 MHz signal) and a Kubo-Toyabe term (quasi zero internal field signal). In addition, dynamical relaxation was accounted for by an exponential damping of the long-time tail of the asymmetry which yields for the total asymmetry $A$

$$A = A_{osc}\exp(-\sigma^2 t^2)\cos(2\pi ft) + A_{KT}G_{KT}(t) + A_{long}\exp(-\lambda t),$$

where $G_{KT}(t)$ is the Kubo-Toyabe function, $\sigma$ represents the damping rate of the 2 MHz signal and $\lambda$ the long-time exponential damping, associated with $T_1$ processes. For $Z$=0.6, since there is no clear evidence for a KT- dip, we rather followed the 2 MHz signal only in 0–5 $\mu$s fits allowing a reliable determination of the frequency and of the damping of the oscillating signal. Whatever the fits performed, the determination of the frequency is quite stable and accurate. The results for the oscillating frequency are presented for both samples in Fig. 9. The absence of variation with $Z$ of both the $T\rightarrow 0$ frequency and its $T$-dependence is one of the most striking results of this study.

The frequency for $Z$=1.2 is found to follow $1.99(2)*(1 -T/T_N)^{0.215(20)}$ MHz with the Néel transition at $T_N$ =395(5)K. The $T$-dependence scales perfectly with that of the ZFNMR as shown in Fig. 9. Above the transition, the ZF asymmetry is slowly relaxing, with a Gaussian KT relaxation rate $\sigma_P$=0.21 $\mu$s$^{-1}$, about twice as large as the nuclear dipole relaxation found in pure $YBa_2Cu_3O_{6+x}$ superconductors. This difference may be partly due to the additional $^1$H nuclear moments. A better accuracy on the transition temperature can be achieved by following the purely paramagnetic response to a small 20 G external field applied perpendicular to the muon polarization. For $Z$=1.2, we find $T_N$=411 K, in fair agreement with the value determined from the frequency variation. As expected from the similarity of the $T$-variations of the internal field detected for $Z$=1.2 and $Z$=0.6, we find a transition temperature for the $Z$=0.6 sample, $T_N$=420(5)K, of the *same order of magnitude* as for $Z$=1.2. The transition is found slightly broader and the frozen fraction smaller as expected from the superconducting to nonsuperconducting fraction determined from susceptibility measurements. Notice the occurrence of a 30% loss of the paramagnetic fraction due to the screening of the external field at the superconducting transition $T_c$=60 K.

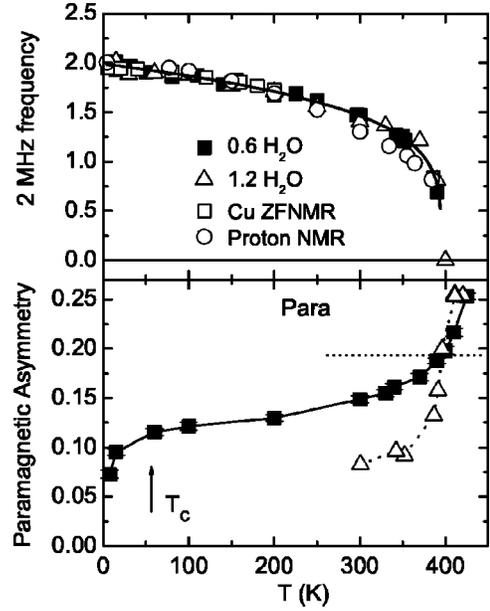

FIG. 9. Top panel: Temperature dependence of the ZF $\mu$SR frequencies in the $Z$=0.6 and 1.2 samples. The $T$-dependence of the Cu ZFNMR local field and the width of the proton line I both scaled to the 2 MHz signal are represented by open squares and circles, respectively. Bottom panel: Plot of the oscillating paramagnetic component obtained in a weak transverse field (ISIS), both for $z$ =0.6 and 1.2. A value of 0.07 is typical for the background obtained for our sample geometry. The "paramagnetic" component corresponds to the unfrozen fraction of the sample.

Finally, in Fig. 10, we plot the Gaussian KT relaxation rate, $\sigma_{KT}$, after subtraction of the nuclear contribution, $\sigma_p$. Above 300 K, an increase in the dynamical relaxation rate obscures the KT dip behavior (dotted curve), and makes extraction of $\sigma_{KT}$ less reliable. $\sigma_{KT}$ decreases faster with $T$ than the frequency of the 2 MHz line (dashed curve), but dies approximately in the same $T$-range, which is an indication in favor of the existence of 2 muon sites probing the *same* AF phase.

In conclusion, our $\mu$SR study shows the existence of a magnetic phase, whose Néel temperature and characteristic internal field does not vary with the water-composition. Two signals can be identified with internal fields of 150 and 10 G, which are likely to originate from the same phase as indicated by the similarity of the temperatures at which the internal fields vanish.

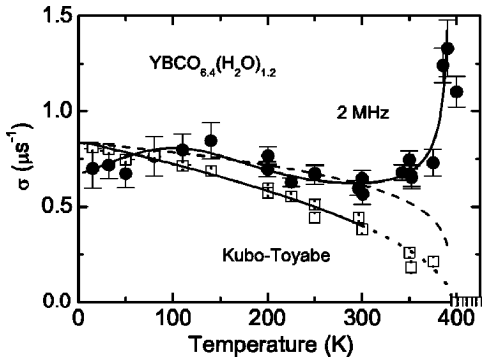

FIG. 10. Gaussian relaxation rate of the 2 MHz signal (filled circles) and Gaussian KT relaxation rates (open squares). The dashed curve is the $T$-dependence of the 2 MHz frequency scaled to the low $T$ value of $\sigma_{KT}$. The dotted curve indicates that the KT dip is not clearly observable in this $T$-range.

### D. Proton NMR

The protons of the inserted water are also good probes of the magnetism of the new phases. We have, therefore, taken proton NMR data on the $YBa_2Cu_3O_{6.4}(H_2O)_{0.6}$ randomly oriented powder sample used for $\mu$SR. The spectrum at low $T$ is almost identical to that published earlier[7,17] and is shown in Fig. 11. It exhibits broad components and can only be obtained by monitoring the spin echo intensity and sweeping the field. The spectrum consists of three lines: Two lines I and II with rectangular shapes and respective widths of 300 and 28 G at low $T$ and a narrow one (III). The features with flat tops are consistent with protons in the presence of an internal magnetic field. Since the sample consists of randomly oriented crystallites, the internal field is oriented at random with respect to the external field. The expected NMR lineshape is "rectangular" with edge steps corresponding to the limiting cases where the internal field is parallel or antiparallel to the external field. Thus the distance between the steps, i.e., the NMR linewidth, is twice the internal field. Therefore we can distinguish two protons sensing local fields of 150 G for site I and 14 G for site II at low $T$, while the narrow line III corresponds to protons which do not sense a significant internal field.

In order to compare the intensities of the lines, one has to correct the spectrum of Fig. 11 from transverse relaxation effects which are found different for the various sites. The transverse relaxation time $T_2$ was found much shorter [$T_2 = 13(1)\mu s$] for the protons of line II than for the those of line I [$T_2 = 111(4)\mu s$]. Extrapolating to zero time delay the spin echo signals of these two lines allowed us to conclude that they have equal intensities within 16% accuracy. The intensity of the central narrow line III which is characterized by a long $T_2 \sim 400$ $\mu s$ was found to be comparatively very small (about 2%) and appears then to correspond to spurious protons located on defect sites, or at the surface of the sample and of the probe coil.

The temperature dependence of the width of the broad line ($\Delta B_B$ in Fig. 11), i.e., the value of the internal magnetic field at the protons site I, has been, after scaling, superimposed on that of the 2 MHz $\mu$SR signal in Fig. 9. One can see that the linewidth scales with the temperature dependence of the 2 MHz $\mu$SR signal and collapses at about 400 K, the same temperature at which the 2 MHz $\mu$SR signal is lost, indicating that the protons of I are located in the same magnetic phase as the muons and confirming its Néel temperature of 400 K. Let us point out that 2 MHz for the muons corresponds exactly to an internal field of 150 G, so that the muons apparently sit on a site identical to that of the proton. This is quite natural as these positively charged particles have the tendency to fill the same positions in the crystal structure. One could then think that muons preferentially sit on proton vacant sites. We could not determine the $T$ dependence of the width of line II, since it is obscured at high $T$ by the spurious proton signal, which also narrows at high $T$ (due to surface water mobility, for example). Therefore, we have no direct evidence that the protons of line II belong to the same magnetic phase as those of line I. However, this is quite likely, as their number is similar to that of protons of line I, and they feel internal fields of the same magnitude as the muons giving a KT term, as pointed out in the previous Section.

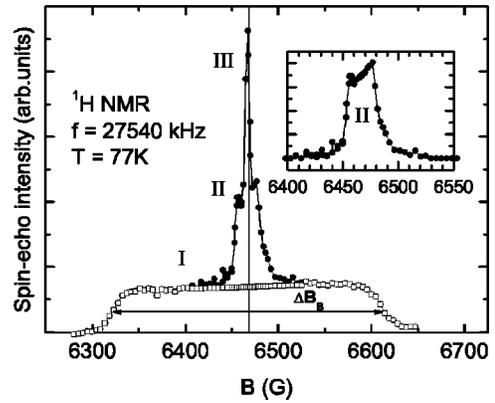

FIG. 11. Proton NMR spectrum for the $YBa_2Cu_3O_{6.4}(H_2O)_{0.62}$ sample taken at 77 K. The decomposition of the spectrum is illustrated. The line shown by open squares corresponds to line I. Line II is obtained by subtraction of line I and of the narrow line III from the total spectrum and is expanded in the inset.

## V. ANALYSIS AND DISCUSSION

Although the presence of magnetic phase(s) in $YBa_2Cu_3O_{6+x}$ samples is evident from our data, the fact that three copper ZFNMR signals and two [1]H and muon signals are detected raises some questions about the assignment of these signals to a single or to multiple phases.

Obviously the proton NMR only shows up two signals with similar intensities, both sensing an internal field. As both $A1$ and $B$ phases deduced from x rays display two distinct proton sites with equal weight, it is then clear that either a single watered phase already occurs for $Z=0.6$, or watered phases are AF with a similar $T_N$ of the order of 400 K and similar internal fields on the proton sites. As $T_N = 420$ K is well known to correspond to the characteristic temperature for the AF of weakly coupled undoped $CuO_2$ bilayers,[18] we might expect that both $A1$ and $B$ phases considered are indeed AF.

But one still needs to understand why we observe three different copper sites detected by ZFNMR. Let us first consider the case where only the single well ordered $B$ phase in Fig. 1 is present in the sample.

### A. Are all the Cu ZFNMR signals due to the $B$ phase?

One can easily distinguish three types of copper ions in this phase, those numbered as 1, 2, and 3 in Fig. 1, so it would be natural to expect three ZFNMR signals with equal intensities. Of course the AF order which is established by insertion of water in the structure involves an ordering of the $Cu^{2+}(2)$ moments. But $Cu(1)$, which is known to have a valence of $2^+$ in filled chains,[19] could order magnetically as well and exhibit a ZFNMR spectrum. This is quite plausible inasmuch as the $Cu(1)$ NQR signals disappear in the water-loaded samples. As one can see in Fig. 6, the intensities of the high- and low-frequency parts of the spectrum are almost equal. That of the central part is substantially smaller and increases faster than others at high water uptake. The high-and low-frequency parts of the spectrum can likely be assigned to the $CuO_2$ bilayer copper sites, while the central ZFNMR spectrum would correspond to the chain copper signal.

In the "watered" samples, holes are localized in the chains by transforming $Cu^+$ into $Cu^{2+}$, which totally suppresses charge carriers in the plane and restores an AF state in the $CuO_2$ bilayers. However, the watered chains certainly modify the geometrical structure of the bilayer with respect to that of the $YBa_2Cu_3O_6$ compound. This can be at the origin of the modification of the on-site and/or supertransferred hyperfine interactions of $Cu(2)$ ions. For example, the watered chains might change the Cu-O bonding angles so that one bonding angle increases up to almost $180°$ and the other one decreases. This is quite possible in this phase as two hydrogens occupy asymmetric positions with respect to the $CuO_2$ planes of the bilayer. The value of the supertransferred hyperfine magnetic field on $Cu(2)$ is known to depend on the bonding angle, being maximal for $180°$ and zero for $90°$. Thus the hyperfine magnetic field on the $Cu(2)$ nuclei of two planes of the bilayer might become quite different which would explain the occurrence of the low-frequency and high-frequency parts of ZFNMR spectrum.

As for the central part of the spectrum it could be assigned to the magnetically ordered copper in the chains. Since $Cu(1)$ in phase $B$ directly neighbors the "defective" chain, its surrounding might be rather asymmetric. The non-zero value of the asymmetry parameter $\eta \approx 0.5$ and/or the angle $\Theta \approx 55°$ between the internal magnetic field and the principal axis of the EFG tensor characteristic for the central part of the spectrum would support this assignment.

*All the local magnetic observations: Three copper signals, two proton and $\mu$SR signals would support then the formation of this unique $B$ magnetic phase.* However, the slowest increase of the $Cu(1)$ signal with water uptake $Z$ would suggest that *the AF ordering of the chain copper is not perfect and only improves with increasing $Z$.*

Let us recall that the studies of magnetic impurities substituted in the chains in $YBa_2Cu_3O_6$ has allowed to evidence that the presence of magnetic moments in the chains fixes the actual 3D magnetic ordering of the AF $Cu(2)$ bilayers in the $c$ direction. The interaction (ferro or AF) between the plane Cu moments and the impurity moment yields a ferromagnetic ordering of the adjacent bilayers in the AFII phase.[12] Here the geometry of $Cu^{2+}(2)$ orbitals with respect to $Cu^{2+}(1)$ ones, would result in a ferromagnetic coupling of their magnetic moments from the semiempirical Goodenough-Kanamori rules.[20] One therefore expects in the $B$ phase not only a freezing of the $Cu(1)$ moments but the occurrence of an AFII phase for the $Cu(2)$ moments. As recalled hereabove, in this phase of the $YBa_2Cu_3O_6$ parent compound it has been clearly shown that the supertransferred magnetic field from $Cu^{2+}(2)$ on $Cu(1)$ nuclei is only about 1 kG. If the transferred field from $Cu^{2+}(2)$ to $Cu^{2+}(1)$ is as small for the water-loaded compound, the hyperfine field of 85 kG at $Cu(1)$ nuclei site is produced mainly by the on-site interaction and the two antiferromagnetically coupled neighbors. The estimates of the hyperfine interaction in the filled $Cu^{2+}(1)O$ chains of $YBa_2Cu_3O_7$ give $A_\perp = 30$ kG/$\mu_B$, $B = 55$ kG/$\mu_B$,[21] and one needs a value of the magnetic moment at the $Cu^{2+}(1)$ site to be about 1 $\mu_B$ to explain the hyperfine field of 85 kG. Of course, a modification of $A_\perp$ and $B$ by the presence of hydrogen certainly occurs as well and might yield a smaller value of the local moment on the $Cu^{2+}(1)$ site.

*One can as well wonder whether the ordering of the* $Cu(1)$ *could affect the spectra of the* $Cu(2)$ *ZFNMR?* Experimentally, with increasing $Z$, the positions of the lines of the high-frequency part of the ZFNMR spectrum remains totally unchanged, but the low-frequency part is slightly shifted (by about 1 MHz) towards low frequencies (Fig. 4, inset). A slight splitting of the lines of this part of the spectrum is already seen for low water uptake. These changes might then be explained by the ordering of the $Cu(1)$ moments with a hyperfine field of about 1 kG supertransferred from $Cu(1)_3$ on $Cu(2)_2$ [but not on $Cu(2)_1$]. Such a magnitude of the supertransferred fields would agree with that known from $Cu(2)$ to $Cu^+(1)$ in $YBa_2Cu_3O_6$. Furthermore, the supertransferred magnetic field from $Cu^{2+}(1)$ should have the sign of the on-site hyperfine field for $Cu(2)$. Since the local magnetic field at the $Cu(2)$ nuclei site is dominated by the transferred field from the four antiferromagnetically coupled neighbors (30 kG each for $YBa_2Cu_3O_6$) which is opposite to the onsite one, the ferromagnetic coupling with $Cu(1)$ should reduce the value of the magnetic field at $Cu(2)$ nuclei, which agrees with the experimental observation. *Such an effect would allow us then to assign the high-frequency and low-frequency ZFNMR spectra respectively to* $Cu(2)_1$ *and* $Cu(2)_2$.

*Is this scenario supported by proton NMR and $\mu$SR data?* Usually the muon senses mainly dipolar fields and occasionally a hyperfine coupling if it stays very near from a magnetic site. Here the fact that the muon and the proton sense the same field suggests that the dominant coupling to their neighbors is dipolar, as the contact hyperfine couplings of the two species should a priori differ. So for consistency, we need to understand whether we can find sites with dipolar magnetic fields of 14 and 150 G in a fully ordered lattice of electronic magnetic moments. Proton locations should be about 1 Å away from the lattice oxygens, which is the typical length of the OH⁻ bond. In high $T_c$ cuprates, the $\mu^+$ is also known to bond to an oxygen to form a $\sim$1 Å O$\mu^-$ bond. The 4 MHz line which is commonly detected in pure or doped AF YBa$_2$Cu$_3$O$_{6+x}$ corresponds to muons near the apical oxygen O(4).[22]

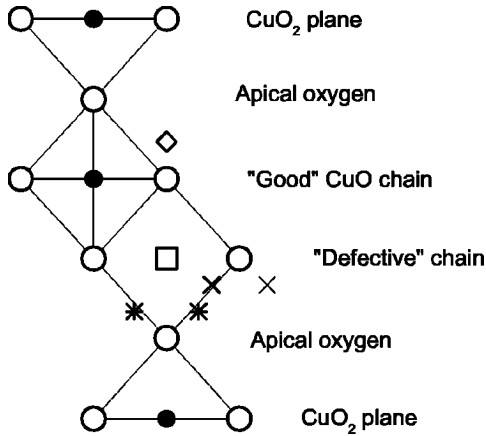

FIG. 12. Possible locations of hydrogen or muon sites sensing a 150 G internal field, obtained from contour plots of the dipolar field calculated with different magnetic structures: diamonds—both Cu(1) and Cu(2) magnetic moments are ordered and directed perpendicular to the chains; crosses and stars—both Cu(1) and Cu(2) magnetic moments are ordered and directed along the chains; at sites marked by stars the internal magnetic field is 150 G (within few Gauss) for both ordered and disordered Cu(1) moments.

So, in the present case, we performed calculations of the maps of the dipolar fields for phase $B$ with lattice parameters $a = b = 0.39$ nm, $c = 2.724$ nm. We took magnetic moments of $0.65\mu_B$ on both Cu(2) and Cu(1), directed perpendicular to the CuO chains, with the stacking sequence of the AFII phase. We find that a dipolar field of 150 G occurs between the CuO chains without Cu vacancies and the CuO$_2$ plane on a site 1 Å far from the chain oxygen (diamonds in Fig. 12). Two quite plausible sites for muons and protons with a 150 G field are also found if the Cu(1) and Cu(2) magnetic moments are taken parallel to the CuO chains (crosses and stars in Fig. 12). For one of them (stars) the field sensed is independent of the Cu(1) magnetization. Let us point out that a 2.2 MHz $\mu$SR signal has been reported[23] in the case of PrBa$_2$Cu$_3$O$_7$, which has an AF phase similar to YBa$_2$Cu$_3$O$_6$. The closeness of this field value to that found in watered YBCO$_{6.5}$ might be associated with an analogous localization of the muon in these phases. In that case the site near the apical oxygen (stars) which does not sense the Cu(1) magnetization would be consistent with both results. Indeed as the Cu chains are filled with O$^{2-}$ in PrBa$_2$Cu$_3$O$_7$ the localization of the muon around the O(4) site might indeed slightly differ from that obtained in the empty chain compound YBCO$_6$.

The calculations for magnetic moments perpendicular to the chains also give dipolar magnetic fields of about 14 G at the site of the missing Cu(1) ions in the chains, which is a very natural location for hydrogen to ensure charge balance. If the Cu(2) magnetic moments are directed along the chain direction, the calculated field at this site is also low (about 6 G) and does not depend on the Cu(1) magnetization. So this site is a very good candidate for the proton and muon site location whatever the actual AF ordering.

### B. What about the other phases?

Although we did not perform systematic x-ray studies of our samples with different water content, published x-ray data suggest that different phases such as $A1$ and $A$ appear as well. Phase $A1$ should occur at low $Z$ and should disappear progressively for $Z > 0.5$. It corresponds to 2 inequivalent bilayer copper sites, analogous in their local structure to those of the $B$ structure, but with weights 3:1. Such a phase corresponds in fact to the formation of a single cell of phase $B$ intercalated within a cell of the starting material. The undoped layers could as well be ordered magnetically through the metallic layer, and one might observe then identical contributions to the low-frequency and high-frequency Cu(2) signals as in phase $B$, probably without ordering of the chain copper. $T_N$ being only weakly affected by the interbilayer coupling,[11,18] the ZFNMR signals might be unaffected by the occurrence of this mixture of phases except the absence of $A1$ in the Cu(1) ZFNMR signal, and the absence of the supertransferred field induced by the Cu(1). So the progressive transformation of phase $A1$ into phase $B$ with increasing $z$ would be in perfect agreement with our ZFNMR observations.

The only question which seems *a priori* incompatible with this scenario is the fact that the proton (muon) internal fields are identical for $Z = 0.6$ and $Z = 1.2$. This would imply that the magnetic ordering of the Cu(1) moments does not modify significantly the dipolar field on the proton and muon sites! However, we do find that such a situation might occur *if the* Cu(1) *and* Cu(2) *moments are aligned in the direction of the chains*. Indeed, at the location at which a 150 G field occurs, that is in the places shown by stars in Fig. 12 which are 1 Å away from the apical oxygen sitting near the defective chain, the dipole contribution of the Cu(1) moments is very small (few Gauss), so that the proton NMR and $\mu$SR should be little affected by the Cu(1) ordering.

Of course, if this is not the actual orientation of the internal field, we are led to consider as an alternative possibility that the central ZFNMR is not that of ordered Cu(1). Then the central part of the ZFNMR spectrum could come from a phase different from phase $B$ (from the products of sample decomposition, for example). In this case the analysis of the low-frequency and high-frequency Cu ZFNMR spectra as due to two Cu(2) sites with different neighborhood would remain the same. There are many sites with dipolar fields of 14 and 150 G in this case, so it is easy to explain both $\mu$SR and proton NMR results. In any case Cu(1) is always 2$^+$ and would exhibit a magnetic moment either paramagnetic or in a frozen disordered state. In the latter case the loss of the NQR spectra of the chain copper is expected. In the former, the fluctuations of the local moment of $Cu(1)^{2+}$ could as well induce a fast spin lattice relaxation of the Cu nuclear spins which would prevent the detection of the Cu(1) NQR signal. Protons might not be present in the phase giving the central part of the ZFNMR spectrum.

However, the main argument against attributing the central Cu NMR signal to a phase distinct from B is that we did not detect a specific $\mu$SR signal which could be attributed to such an ordered magnetic phase.

### C. Samples packed in Stycast

The changes of the copper NQR spectra for the $YBa_2Cu_3O_{6+x}$ ($x=0.4-0.7$) samples packed in Stycast and annealed at 200°C in air are in general the same as those for the free-powder annealed samples. During the first two hours of annealing of the $x=0.5$ powder packed in Stycast, the NQR spectrum intensity decreased by 50% and then slowly reduced to 20% of the initial intensity after 80 hours of annealing.

The ZFNMR spectrum after 80 hours annealing is similar to that of the $Z=1.2$ free-powder annealed spectrum (Fig. 1 in Ref. 4), the only difference being the less resolved spectra at 45–96 and 96–140 MHz. The superconducting volume fraction decreased during the annealing at the same rate as the copper NQR intensity, but slower than for free powders. A 80% decrease of the NQR signal was achieved after 80 hours. Similar low intensity ZFNMR spectra were observed also in Stycast-packed samples stored for few years at room temperature. This indicates that minute amounts of the samples are already transformed after long time reaction with Stycast at room $T$.

In order to check whether only $YBa_2Cu_3O_{6+x}$ compounds with semi-empty chains, i.e., those with $x<1$, are subject to topotactic reaction with water vapor or Stycast decomposition products, one $YBa_2Cu_3O_7$ sample packed in Stycast was annealed at 200°C for 5 hours. The superconducting volume fraction decreased by about 30% after such treatment, as was reported earlier.[4] At the same time, a magnetic phase appeared in the sample with a copper ZFNMR spectrum with slightly lower hyperfine magnetic fields than watered $YBa_2Cu_3O_{6.5}$ samples.

Since Stycast somehow protects the samples from air moisture, the sample degradation most likely occurs through insertion of the products of Stycast decomposition. It is known that the polymer chains of most types of epoxy resin contain OH-groups which apparently can readily tear off the polymer chain, especially shortly after polymerization, and react with the superconductor material. Elevated temperatures promote fast polymerization and thus stop the emission of OH-groups in a shortwhile. The result of the reaction with OH-groups is most likely the same as that with water - appearance of an additional chain in the crystal structure and conversion of the 123 compound into the nonsuperconducting pseudo-248. The faster decrease of the superconducting volume fraction in samples reacting with water than with Stycast could be explained by the extra hydrogen in the $H_2O$ molecule which binds a hole in the $CuO_2$ plane[24] and thus promotes a faster decrease of the superconductivity.

### VI. CONCLUSIONS

Our copper NQR/ZFNMR, $\mu$SR and proton NMR studies of the $YBa_2Cu_3O_{6+x}$ compounds reacted with water vapor give straightforward evidence that water insertion proceeds dominantly through the empty Cu(1) chains. The most ordered regions of the crystallites are most easily transformed. The water insertion is very slow at room temperature, but takes place in a few days at 100–200°C. The products of the reaction are non-superconducting antiferromagnetic phases in which the bilayer order at $T_N \approx 400$ K. The two copper sites of the magnetic bilayers display quite different ZFNMR signals with internal fields of 62 kG and 101 kG. A third ZFNMR signal with an internal field of 85 kG appears already for low water uptake and its intensity increases markedly for high water uptake.

We have shown above that phase $B$ would explain all the results for the highest water content with or without deciding unambiguously whether the 85 kG Cu ZFNMR spectrum is due to chain ordered Cu(1) moments. It could be eventually attributed to a parasitic phase. However, an assignment of this central part of the ZFNMR spectrum to Cu(1) seems quite reasonable if the moments are aligned along the chain axis. In that case our data are compatible with the appearance of a mixture of phases $A1$ and $B$ at low water content with a progressive ordering of the Cu(1) moments and the corresponding transformation from the AFI to the AFII 3D magnetic order with increasing water content. Such a scenario explains both the slight frequency shift of the low-frequency Cu ZFNMR spectra associated with the progressive increase of AFII phase content, and the absence of detectable modification of the $\mu$SR and proton NMR spectra.

$YBa_2Cu_3O_{6+x}$ samples are very sensitive to air moisture, and even samples packed in paraffin underwent a transformation of a large part of the sample after 6 years storage in the ambient atmosphere.[3] Samples (both $YBa_2Cu_3O_{6.4-6.7}$ and $YBa_2Cu_3O_7$) packed in Stycast heated at 200°C, for which pure Stycast does not fully collapse, still undergo some epoxy decomposition and yields the formation of the same AF compound (about 20% after 80 hours for $YBa_2Cu_3O_{6.5}$, with a grain size of about 20 microns). Traces of such a reaction could be found even in samples packed in Stycast and stored at room temperature for few years. It is then quite clear that such effects should be considered quite seriously and avoided in experiments attempting to resolve tiny effects in these materials, such as for instance those performed by neutron scattering.

### ACKNOWLEDGMENTS

This study was supported in part by the Russian Foundation for Basic Research, under Project 03-02-16550, and by the Russian Ministry of Industry and Science, under Project 98014-4. P.M., A.M.F., and J.B. thank the PSI and ISIS facilities for their kind hospitality and support of the project. The $\mu$SR experiments at ISIS were supported by the EU-TMR program. The financial support of CRDF under Grant No. BRHE REC-007 is also appreciated by A.V.D., A.V.E., I.R.M., and A.V.S.